\documentstyle[rotating,epsfig,amsmath,amssymb]{elsart}

\def\be{\begin{equation}}
\def\ee{\end{equation}}
\def\bea{\begin{eqnarray}}
\def\eea{\end{eqnarray}}
\newcommand{\beq}{\begin{equation}}
\newcommand{\eeq}{\end{equation}}

\begin{document}
\def\theaddress{\arabic{address}}

%%%%%%%%%%%%%%%%%%%%%%%%%%%%%%%%%%%%%%%%%%%%%%%%%%%%%%%%%%%%%%%%%%%%%%%%%%%%%

\begin{frontmatter}

\title{The HERMES Dual-Radiator \\ Ring Imaging \v{C}erenkov Detector}

\collab{\em The HERMES RICH group}

\author[yer]{N.~Akopov}, 
\author[desyz]{E.C.~Aschenauer}, 
\author[anl]{K.~Bailey}, 
\author[tokyo]{S.~Bernreuther},
\author[fras]{N.~Bianchi}, 
\author[fras]{G.~P.~Capitani}, 
\author[calt]{P.~Carter}, 
\author[rome]{E.~Cisbani}, 
\author[bari]{R.~De~Leo},
\author[fras]{E.~De Sanctis}, 
\author[anl]{D.~De~Schepper}, 
\author[desyz]{V.~Djordjadze}, 
\author[calt]{B.W.~Filippone}, 
\author[rome]{S.~Frullani}, 
\author[rome]{F.~Garibaldi}, 
\author[anl]{J.-O.~Hansen}, 
\author[gent]{B.~Hommez}, 
\author[rome]{M.~Iodice}, 
\author[anl]{H.~E.~Jackson}$^{,\ast}$
\renewcommand{\thefootnote}{\fnsymbol{footnote}}
\footnotetext[1]{Corresponding author. Tel.:+630-252-4013. e-mail:HAL@anl.gov}
\renewcommand{\thefootnote}{\arabic{footnote}}\hspace{-5mm}, 
\author[desyz]{P.~Jung},
\author[desyz]{R.~Kaiser}, 
\author[tokyo]{J.~Kanesaka},
\author[anl]{R.~Kowalczyk}, 
\author[bari]{L.~Lagamba},
\author[desyz]{A.~Maas}, 
\author[fras]{V.~Muccifora}, 
\author[bari]{E.~Nappi},
\author[desyz]{K.~Negodaeva}, 
\author[desyz]{W.-D.~Nowak},
\author[anl]{T.~O'Connor},  
\author[anl]{T.~G.~O'Neill}, 
\author[anl]{D.~H.~Potterveld}, 
\author[gent]{D.~Ryckbosch}, 
\author[tokyo]{Y.~Sakemi}, 
\author[tokyo]{F.~Sato},
\author[desyz]{A.~Schwind}, 
\author[tokyo]{T.-A.~Shibata},
\author[tokyo]{K.~Suetsugu},
\author[fras]{E.~Thomas}, 
\author[gent]{M.~Tytgat}, 
\author[rome]{G.M.~Urciuoli}, 
\author[gent]{K.~Van~de~Kerckhove}, 
\author[gent]{R.~Van~de~Vyver}, 
\author[tokyo]{S.~Yoneyama}, 
\author[yer]{H.~Zohrabian},
\author[tokyo]{L.~F.~Zhang}
\address[anl]{Physics Division, Argonne National Laboratory, Argonne, 
IL 60439, USA}
\address[bari]{INFN, Sezione di Bari, 70124 Bari, Italy}
\address[calt]{W.K.Kellogg Radiation Lab, California Institute of Technology, \\
Pasadena, CA 91125, USA}
\address[desyz]{DESY Zeuthen, 15738 Zeuthen, Germany}
\address[fras]{INFN, Laboratori Nazionali di 
Frascati, 00044 Frascati, Italy}
\address[gent]{Dept. of Subatomic and Radiation Physics, University 
of Gent,\\ 9000 Gent, Belgium}
\address[rome]{INFN, Sezione Roma1 - Gruppo Sanit\`a, 00161 Roma, Italy}
\address[tokyo]{Department of Physics, Tokyo Institute of Technology, 
Tokyo 152-8551, Japan}
\address[yer]{Yerevan Physics Institute, 375036 Yerevan, Armenia}
\newpage

\begin{abstract}
The construction and use of a dual radiator Ring Imaging \v{C}erenkov (RICH)
detector is described. This instrument was developed for the HERMES
experiment at DESY which emphasises measurements of semi-inclusive
deep-inelastic scattering. It provides particle identification
for pions, kaons, and protons in the momentum range from 2 to 15 GeV,
which is essential to these studies. The instrument  uses two
radiators, ${\mathrm C_4F_{10}}$, a heavy fluorocarbon gas, 
and a wall of silica aerogel tiles. The use of aerogel in a RICH
detector has only recently become possible with the development of
clear, large, homogeneous and hydrophobic aerogel. A lightweight mirror was
constructed using a newly perfected technique to make resin-coated
carbon-fiber surfaces of optical quality. The photon detector consists
of 1934 photomultiplier tubes (PMT) for each detector half, held in a 
soft steel matrix to provide shielding against the residual field of 
the main spectrometer magnet. 
\end{abstract}
\end{frontmatter}

\twocolumn

\section{Introduction}

The HERMES experiment\cite{spectrometer} 
is a study of the spin structure of the nucleon
which emphasizes on an unambiguous
measurement of pion, kaon, and proton semi-inclu\-sive spin 
asymmetries in deep-inelastic scattering (DIS). These asymmetries 
provide information on the flavor dependence of 
polarized structure functions and the sea polarization. 
However, most of the hadrons produced in HERMES~\cite{Cis1} 
lie between 2~and 10~GeV, a region in which it has not been 
feasible to separate pions, kaons, and protons with standard 
particle identification (PID) techniques. 
Ring imaging \v{C}eren\-kov (RICH) and threshold \v{C}erenkov systems using heavy 
gases~\cite{segu94}, such as ${\mathrm C_{4}F_{10}}$, at atmospheric pressure are 
useful only for energies above 10~GeV since the kaon \linebreak threshold for 
\v{C}erenkov  radiation is typically higher than 9~GeV.
Because of substantial multiple scattering and brems-{\linebreak}strahlung, the  
use of a high pressure gas system is not technically feasible in 
HERMES. Clear liquid radiators are only useful for hadron identification below 
roughly 2~GeV because of their 
very low \v{C}erenkov light thresholds and large chromatic dispersion. 

With the recent development\cite{yokogawa,adac} of new clear silica
aerogel with a low index of refraction, this difficult energy region
can now be spanned. Aerogels have 
long been used in threshold \v{C}erenkov counters~\cite{burk}, but their 
use as a radiator in a RICH system is a recent development~\cite{fiel}
which has resulted from the availability of material with excellent optical 
properties. 
This paper reports the successful use, for the first time, of clear aerogel
in combination with a heavy gas, ${\mathrm C_{4}F_{10}}$, in a RICH 
detector. This dual-radiator RICH detector provides clean separation of pions, 
kaons, and protons over most of the kinematic acceptance of the HERMES experiment. 
Such a configuration was first proposed for the planned\linebreak LHCb experiment~\cite{forty}.

The following section~2 describes the general design of the HERMES RICH detector
and the properties of its key components. Section~3 describes briefly
the online monitoring of the system, section~4 the reconstruction and
particle identification and section~5 the necessary software
correction of the mirror alignment. Section~6 finally is dedicated to
the performance of the detector.

\section{Detector Design}

\subsection{Design Requirements}

HERMES is located in the East hall of the HERA storage ring complex. 
The spectrometer is split into two halves above and below a central
horizontal magnetic shielding 
plate through which the HERA beams pass~\cite{spectrometer}. 
As a result, the RICH detector also consists of two symmetric RICH modules, 
the top and the bottom RICH sections.  
The RICH is positioned between the two rear tracking chambers, so that
the free space is only 1.25~m deep. 
The outer body and the gas control system were taken unchanged from the
previous threshold \linebreak \v{C}erenkov counters.

\begin{figure}[t]
\epsfxsize=3.0in \epsfbox{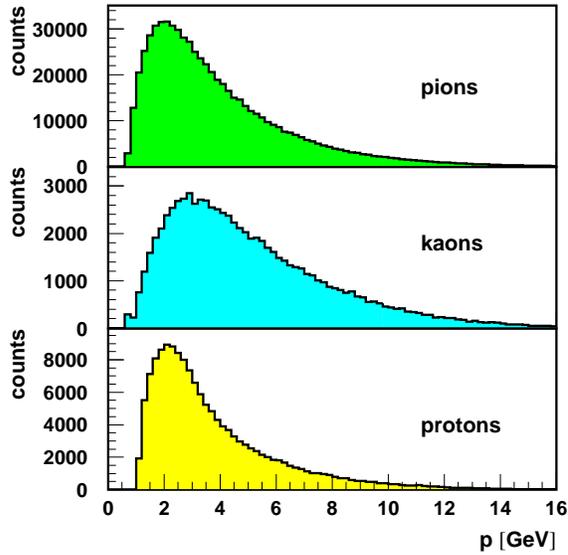}
\caption{Monte Carlo hadron momentum spectra within the HERMES acceptance.}
\label{fig:hadron_momenta}
\end{figure}

Figure~\ref{fig:hadron_momenta} shows the hadron momentum 
spectra in the HERMES acceptance obtained from a GEANT based Monte Carlo (MC)
simulation of the deep inelastic scattering of 27.5 GeV positrons off
a hydrogen target. The fall-off 
of the spectra at low momenta is due to the field of the spectrometer magnet, which
severely limits the acceptance at lower momenta.
About 95\% of all hadrons in the acceptance are found in the range of 2.0 to 15.0 GeV.
This defines the momentum range over which clear particle
identification should be provided.

\begin{figure}[h]
\begin{center}
\epsfysize=3.in \leavevmode \epsfbox{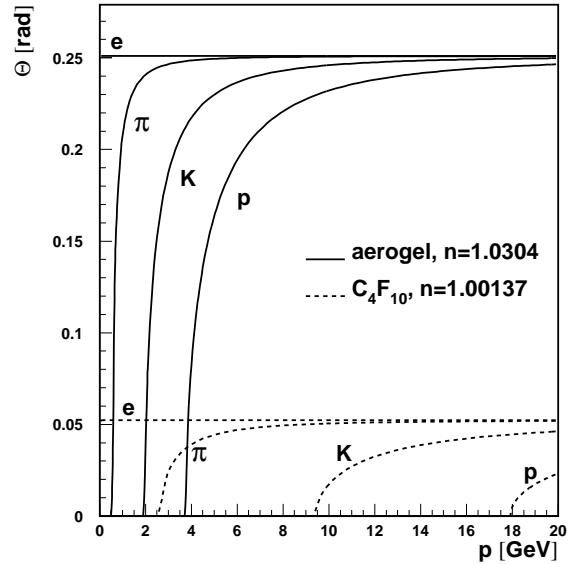}
\end{center}
\begin{center}\begin{minipage}{3.0in}
\caption[]{The \v{C}erenkov angle $\theta$ versus hadron momentum for the
aerogel and ${\mathrm C_{4}F_{10}}$ gas radiators.
\label{cerangles}}
\end{minipage}\end{center}
\end{figure}

The low end of this range determines the index of refraction necessary for the
aerogel. A value of n($\lambda$=633~nm)=1.03 was chosen since it leads to
a kaon threshold of 2 GeV. The \v{C}erenkov angles produced by the combination
of this aerogel and the heavy gas (${\mathrm C_{4}F_{10}}$) for pions, 
kaons and protons are plotted in figure~\ref{cerangles} as a function 
of particle momentum.
The corresponding threshold momenta are listed in table~\ref{table:thresholds}.
All pion momenta within the spectrometer acceptance are above the pion
threshold momentum for aerogel of 0.6~GeV, 90\% of the kaon and 78\%
of the proton momenta are above the kaon threshold of 2.0~GeV. 

\begin{table}[h]\centering
\begin{tabular}{|c||c|c|}
\hline
~ & aerogel & ${\mathrm C_4F_{10}}$ \\
\hline
n & 1.0304 & 1.00137 \\
$\beta_t\gamma_t$ & 4.03 & 19.10 \\
$\pi$ & 0.6 GeV & 2.7 GeV \\
$K$   & 2.0 GeV & 9.4 GeV \\
$p$   & 3.8 GeV & 17.9 GeV \\
\hline
\end{tabular}
\vspace*{0.2cm}
\caption[]{\v{C}erenkov light thresholds for pions, kaons and protons.
The index of refraction n is given at 633 nm, $\beta_t=1/n$ is the threshold
velocity and $\gamma_t=1/\sqrt{1-\beta_t^2}$.}
\label{table:thresholds}
\end{table}

The high end of the momentum range fixes the number of photons
that must be detected for full hadron separation. The parameter
to be considered is $p_{max}$, the maximum separation momentum~\cite{bib2}. 
This is defined as the maximum momentum for which the
average photon emission angle of two particle types (with
masses $m_1$ and $m_2$) is separated by a number of standard deviations 
$n_{\sigma} $ :
\begin{equation}
p_{max} = \sqrt{ \frac{m^2_2 - m^2_1}{2 k_f n_{\sigma}}}
\label{pmax}
\end{equation}
where $k_f = \tan \theta \cdot\sigma _{\theta} / \sqrt{N}$ is the RICH detector
constant, $N$ is the number of separately detected photons,
$\theta$ is the \v{C}erenkov angle and $\sigma _{\theta}$ the standard deviation 
of the reconstructed photon angle distribution. 
In the design, $n_{\sigma}=4.652$ was chosen, as it corresponds 
to a misidentification of the particle in 1\% of the cases, assuming equal 
fluxes for the two particle types, an average detector response (in 
yield and resolution) and no background.

Assuming $\sigma _{\theta}$ to be 7 mrad (see tables \ref{table:delta} 
and \ref{table:delta2}) 
it follows from (\ref{pmax})
that $p_{max}(\pi,K)$ = 15 GeV requires $N$ for the gas to be 12. This
requirement leads to the design values for $p_{max}$ given in table
\ref{table:pmax}. In this estimate it was assumed that the number
of separately detected photons from the aerogel is 10. 
\begin{table}[b]\centering
\begin{tabular}{|c||c|c|}
\hline
~ & aerogel & ${\mathrm C_4F_{10}}$ \\
\hline
$k_f$ & $5.46\cdot10^{-4}$ & $1.07\cdot10^{-4}$   \\
$p_{max}(\pi/K)$ & ~6.7 GeV & 15.0 GeV \\
$p_{max}(K/p)$   & 11.2 GeV & 25.3 GeV \\ 
\hline
\end{tabular}
\vspace*{0.2cm}
\caption{Maximum separation momenta $p_{max}$.}
\label{table:pmax}
\end{table}
\begin{figure}[b]
\includegraphics[width=3.0in]{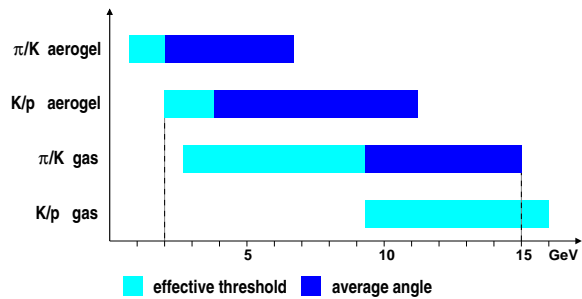}
\caption{Momentum ranges for hadron separation in aerogel and
${\mathrm C_4F_{10}}$. Between the dashed lines the hadrons
can be separated.}
\label{fig:pmax}
\end{figure}
Figure~\ref{fig:pmax} illustrates the overlap
between the momentum regions for both radiators. The lightly shaded
region indicates where the particle can be identified based on whether
or not a ring is present at all. In this region the detector acts like
a threshold \v{C}erenkov. In the darkly shaded region the identification
is based on the average reconstructed angles. The plot considers each
radiator separately, but the PID algorithms will combine the information
from the two.
The momentum region for
which the identification of pions, kaons and protons is possible is limited
by the kaon threshold momentum for aerogel at 2.0 GeV and by the 
maximum separation
momentum for $\pi$/K separation in ${\mathrm C_4F_{10}}$ at 15.0 GeV.

\subsection{General Design Parameters}

\begin{figure}[b]
\vspace*{-0.3cm}
\begin{center}
\epsfysize=3.0in \leavevmode \epsfbox{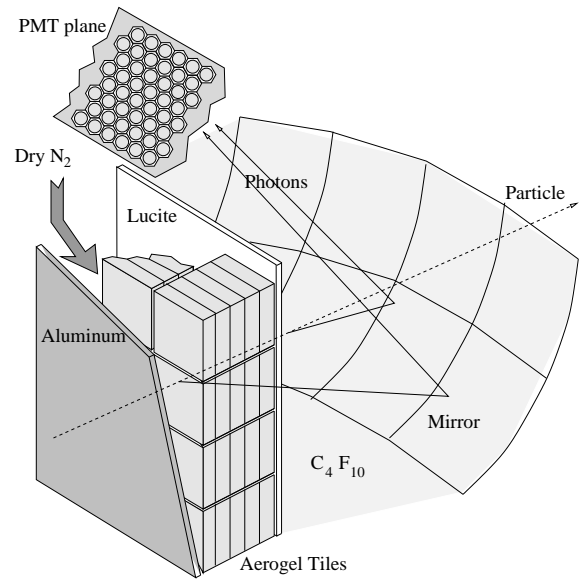}
\end{center}
\begin{center}\begin{minipage}{2.8in}
\caption[]{Basic geometry and radiator configuration for the HERMES dual 
radiator RICH (not to scale).
\label{RICHview}}
\end{minipage}\end{center}
\end{figure}

\enlargethispage{\baselineskip}
The geometry which was adopted for the \v{C}erenkov radiators and ring imaging
systems is shown in figure~\ref{RICHview}~\cite{cisbani}. 
The body of the counter is constructed of aluminum, with entrance and exit 
windows made of 1 mm thick aluminum. 
The volume of each half is approximately 4000~l.
The size of the entrance window is 187.7~cm by 46.4~cm and the exit 
window 257.0~cm by 59.0~cm.

\begin{figure*}[t]
\begin{center}
\epsfysize=3.0in \leavevmode \epsfbox{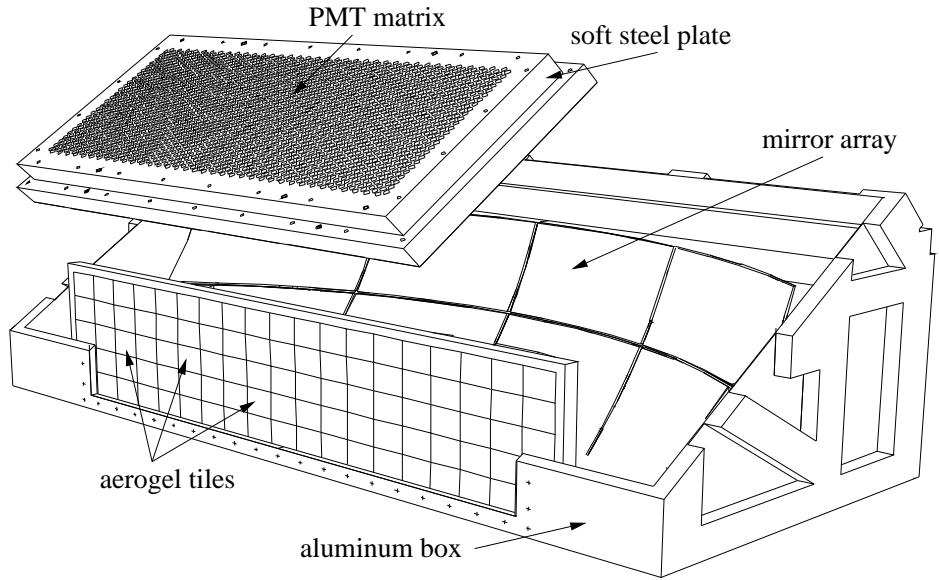}
\caption[]{A cutaway schematic view of the (top) RICH counter. }
\label{3dcutbw}
\end{center}
\end{figure*} 
\enlargethispage{\baselineskip}
A gas control system recirculates the radiator gas through the main volume,
keeping the gas at a slight overpressure with respect to
atmosphere. The aerogel radiator is an 
assembly of tiles configured to fill 
the entrance of the detector with an aerogel thickness of 5.5~cm. The 
unoccupied 
volume of the detector behind the aerogel is filled with the gas radiator, 
${\mathrm C_{4}F_{10}}$. A spherical mirror array located at the rear of the 
radiator box images the \v{C}erenkov light cones onto 
a focal surface located \linebreak above (below) the active volume. 

The radius of curvature of the mirror array is 2.20~m. 
It was chosen to give a focal 
surface location in the accessible region above (below) the forward region of the 
radiator boxes and to provide a detector plane of tractable dimensions.
The optical axis of the array, the perpendicular to the mirror surface at
the center of the array, is inclined at an angle of 26 degrees to the
horizon. The photon detector is located outside of the mirror optical axis
with its axis inclined at an angle of 40 degrees to the horizon so as to
intercept the mirror surface at a distance of 90 cm.
The focal length of the mirror is 110 cm. 
The boxes are fitted with gas connections and pressure regulators which provide a continuous 
controlled flow of recirculating gas. An open section of one of the RICH counters is shown in 
figure~\ref{3dcutbw}.

The size of a useful detector
surface was evaluated by an MC simulation which included an early
version of the RICH  geometry described above. 
The simulation showed that 
95\% of the centers of the rings and 90\% of all the photons are contained in 
a planar surface 60~cm high and 120~cm wide (0.72~m$^2$ surface area).
These dimensions were used as lower limits
in the final design of the photon detector.

The inner walls of the box are blackened to reduce wall reflections.
An array of green light-emitting diodes (LED) 
is installed to provide test
and calibration pulses for the photon detector. They are
located on the face of the mirror, so as to provide an 
approximately uniform illumination of the \linebreak photon detector surface. 

As explained below, most of the useful photons contained in rings
emitted from the aerogel are
in the visible light region. A good choice for the photon detector is then
an array of photomultiplier tubes (PMTs). In the present design 3/4 inch
tubes were used, leading to a pixel size of 23.3 mm. A total of 1934 PMTs
is necessary to cover the photon detector area.

Table~\ref{table:delta} lists estimates for
the different contributions to the single photon resolution.
$\Delta {\theta}_{em}$ results from the uncertainty in the emission
vertex along the track in the radiator.
The detector granularity determines $\Delta {\theta}_{pix}$, which follows
from the size of the ``pixels''.
These two contributions affect \linebreak both the aerogel and the gas angles.
The chromatic aberration $\Delta {\theta}_{chr}$ derives from 
the variation of the index of 
refraction with respect to the wavelength and is only important
for the resolution in the aerogel.
These quantities do not add  quadratically
and are only listed separately to indicate their relative importance. 
Their combined effect is calculated in the MC and is 
listed as $\Delta{\theta}_{MC}$.
The resolution is dominated by the pixel size, which is determined 
by the size of the PMT and how much material must
surround it to provide adequate magnetic shielding. 

\begin{table}[ht]\centering
\begin{tabular}{|l||c|c|}
\hline
~ & aerogel & ${\mathrm C_4F_{10}}$ \\
\hline
$\Delta {\theta}_{em}$   & 1.8 mrad & 2.2 mrad \\
$\Delta {\theta}_{pix}$  & 5.6 mrad & 5.2 mrad \\
$\Delta {\theta}_{chr}$  & 2.5 mrad &  - \\    
\hline
$\Delta {\theta}_{MC}$ & 7.1 mrad & 7.2 mrad \\ 
\hline  
\end{tabular}
\vspace*{0.2cm}
\caption{Individual contributions to the single photon resolution in
aerogel and ${\mathrm C_4F_{10}}$.}
\label{table:delta}
\end{table}

\subsection{Aerogel Radiator}

New production techniques~\cite{yokogawa,adac} have yiel\-ded 
aerogels with much more uniform and smaller colloidal ${\mathrm SiO_2}$ particle 
structure.  The more uniform granular structure suppresses Rayleigh 
scattering of visible light and greatly enhances its transmission. 
Material with a refractive index in the range 
$1.01-1.10$, suitable for use in the momentum range of interest in HERMES,
is now commercially available from Matsu\-shita Electric 
Works\footnote{Advanced Technology Research Laboratory,
1048 Kadoma, Kadoma-shi, Osaka-fu, Japan 571,
contact: M. Yokoyama, email: yokoyama@crl.mew.co.jp} 
in the form of tiles of average dimensions 114 by 114 by 
11.3 mm$^3$. 
The optical properties of the aerogel tiles have been studied in detail using
a sample of 200 tiles. The transmittance was measured for each tile for the
wavelength range of 200-900 nm. Typical values for individual tiles at 
400 nm are around
0.67. The transmittance $T$ was parameterized as a function of wavelength 
$\lambda$ and aerogel thickness $t$ by the Hunt formula~\cite{del1}:
\begin{equation}
T = A \cdot e^{\frac{-C \cdot t}{\lambda^4}} 
\end{equation}
The measurements can be described well with a value 
of $C \cdot t$~=~0.0094~$\mu$m$^4$ and $A=0.964$.
More details on the optical properties of the
aerogel tiles can be found in~\cite{zhang,optical,del1,Car1}.

Out of 1680 tiles produced in a single production batch, 
1020 tiles were judged to be usable for the HERMES RICH. The selection 
was based on the measured refractive index and the sizes of the individual
tiles~\cite{zhang}. 
Of the 1020 tiles, 850 were actually used. For those tiles, 
the average index of refraction is 1.0304, with a spread of 
$3.6\cdot10^{-4}$~\cite{optical,zhang}. The tiles are stacked in a container
consisting of an aluminum frame with a 1~mm aluminum entrance window and a 
3.2~mm UVT-lucite exit window. 
To guard against possible de\-gradation of the aerogel by the
${\mathrm C_{4}F_{10}}$ environment, the aerogel container is sealed 
gas tight and dry nitrogen is continuously circulated \linebreak through the box at a 
slow rate.

The aerogel tiles are stacked in 5 layers, with 5 horizontal rows, and
17 vertical co\-lumns as required to span the spectrometer acceptance. 
Black plastic spacers of appropriate thicknesses between 
the aluminum frame and the tiles prevent them from shifting while 
the radiator is moved. 
The aerogel wall thickness was chosen as the optimal point for maximizing 
the unscattered light yield relative to scattered background 
photons~\cite{del1,Car1}. The tiles
are stacked according to  their measured refractive
index, thickness, and surface quality. To achieve
the best ring resolution, tiles with similar refractive indices are placed
together in the same stack, so that particles passing through the radiator
emit photons with very similar \v{C}erenkov angles. 
Opaque black sheets of tedlar between the aerogel stacks  
reduce distortions by absorbing photons that cross stack boundaries. 
\begin{figure*}[t]
\begin{center}
\epsfysize=2.4in \epsfbox{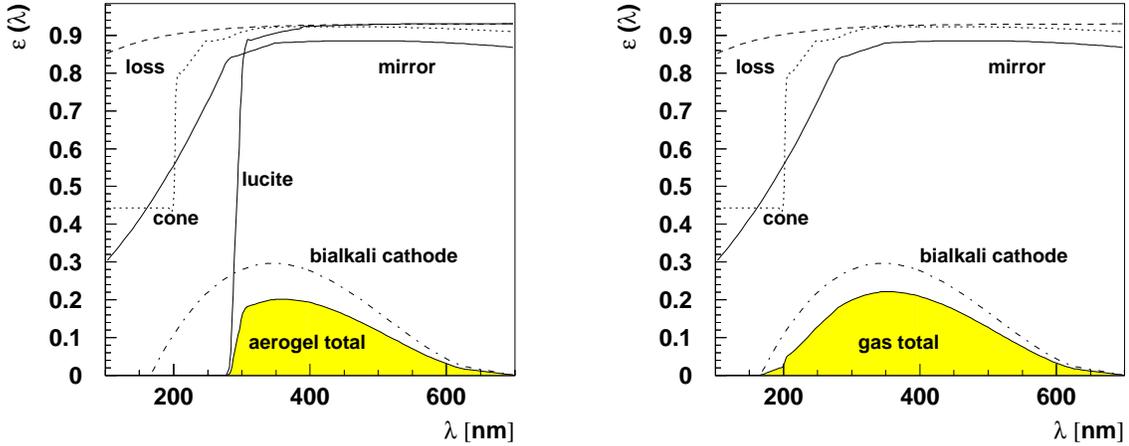}
\vspace*{-0.3cm}
\caption{Efficiencies for transmission and detection of aerogel (left) and
gas photons (right). The included contributions are the mirror
reflectivity (mirror), the PMT cone efficiency (cone), the effect of
the air gap between PMT and quartz window (loss), the light
transmission through lucite and the quantum 
efficiency of the PMT bialkali cathode. The sharp edge in the cone 
efficiency results from the sharp cutoff in reflectivity at 200 nm.
The shaded histogram is the total efficiency to detect an emitted photon.}
\label{fig:effies}
\end{center}
\end{figure*} 

The choice of the material for the exit window was driven by its 
transmission properties.
Due to the proportionality to $\lambda^{-4}$ of the Rayleigh scattering 
cross section, the low wavelength end of the light yield spectrum is
dominated by rescattered photons. Since they only contribute to the background,
the exit window material was selected to absorb most of them. 
A good match was found in ultraviolet-transmitting (UVT) lucite, which 
has an absorption cutoff of 290 nm~\cite{Car1}(see figure~\ref{fig:effies}). 
The thickness of the UVT window, 3.2 mm,  was chosen to be as small
as possible while still providing sufficient mechanical stability. 
It contributes less than 1\% of a radiation length.
The \v{C}erenkov light generated in the lucite itself is emitted at such high
angles that it is effectively trapped by total internal reflection for most 
forward tracks.

\subsection{Mirror Array}

The mirror arrays each consist of eight segments that are mounted in 
two rows of four 
(see figure~\ref{3dcutbw}). Both the mirrors
and the mounting frame were fabricated by Composite 
Mirror Applications (Tucson, Arizona, USA)\footnote{
web: http://home.earthlink.net/$\sim$bromeo/\newline
contact:~R.~Romeo,~email:~bromeo@earthlink.net
}.
The backing of the segments is fabricated from a graphite fiber 
composite~\cite{Che1} coated with an epoxy film yielding an optically 
smooth surface. Surface roughness was specified to be less than 5~nm. 
The surface is aluminized to provide a reflectivity above 85\% 
for light in the 300-600~nm range (see figure~\ref{fig:effies}). 
The segments are held on a
carbon-fiber frame with individual adjustable three point mounts. 
The dimensions of each
array are 252.4~cm by 79.4~cm; each weighs less than 13~kg. This guarantees
that the mirrors contribute negligibly (less than 1.0\% of a radiation 
length) to the total radiation length of the detector, which is
dominated by the freon gas (5\%), the aluminum windows (3\%) and the aerogel (2.8\%).

The short time scale of the RICH installation (13 months from approval to
installation) only allowed for a simple alignment procedure
when the mirrors were installed.
A point light source at the nominal radius of curvature of the array was 
imaged on a screen. The mirror segments were then adjusted to minimize the 
size of the reflected image. To neutralize the unexpected flexibility of the
mirrors, their mounting was fine-tuned by adjusting additional 
pressure points where necessary.
A nominal 90\% of the reflected light was 
contained in a spot corresponding in size to the pixel size of 23~mm. 
The actual mirror alignment was determined from the data itself 
(see section \ref{alignment_section}).

\subsection{Photon Detector}

As discussed, the useful photons emitted in the aerogel are largely limited
to visible wavelengths. 
The radiation emitted in the gas does have a component 
extending into the UV region, but since the photon yield from the gas 
is sufficiently high, the detector response in the UV is not crucial. 
Consequently, photon-to-electron conversion by 
photocathodes typical of commercial photomultipliers
provides a simple and robust photon detector technology. 

While photomultipliers with a diameter of 0.5~inch are commercially available, 
their cost and the time of construction for two large 0.5~inch arrays were
unacceptable. Instead, Philips XP1911 photomultipliers  were chosen, 
with a diameter of 18.6~mm (0.75~inch) and a guaranteed minimum active
photocathode diameter of 15~mm~\cite{elke}. This led to a pixel size of 23.3~mm.
$\Delta{\theta}_{pix}$ can be estimated at the center of the detector 
by assuming that both the radiator and the detector are at the focal plane. 
This gives $\Delta{\theta}_{pix}$~=~$s/(4f)=5.3$~mrad (for circular pixels), 
where $f$ is the focal length 
(110~cm) and $s$ is the linear pixel size. However, this is
a simplifying assumption and an MC simulation leads to the values in
table \ref{table:delta}.

The XP1911/UV (green enhanced) PMT was chosen for its broad 
quantum efficiency curve that matches
the \v{C}erenkov light spectrum of the aerogel well. Since the
efficiency extends into the UV, it is also a good choice for the
gas photons, as is clear from figure~\ref{fig:effies}. 
This figure shows the different contributions
to the total photon detection efficiency for both radiators.

An automated PMT testing procedure was used to
select PMTs with a broad plateau. High voltage (HV) plateau curves were
measured for each tube with a 470~nm light emitting diode pulser. In addition,
a low dark current single photoelectron pulse rate was required. After
the installation of the RICH in the HERMES spectrometer a noise rate
of about 1~kHz was reached, corresponding to 1 PMT firing per detector 
half every five events (gate length 100~ns). 
The PMTs were sorted according to their gain
characteristics and their noise level.
With an especially developed low current (0.040 mA at 1300 V) 
high voltage divider it
was possible to supply 32 PMTs from a single HV channel. The
divider also limited the heat generated in the base of the PMTs, 
thereby significantly reducing the cooling requirements of the detector.

\begin{figure}[t]
\includegraphics[width=3.0in]{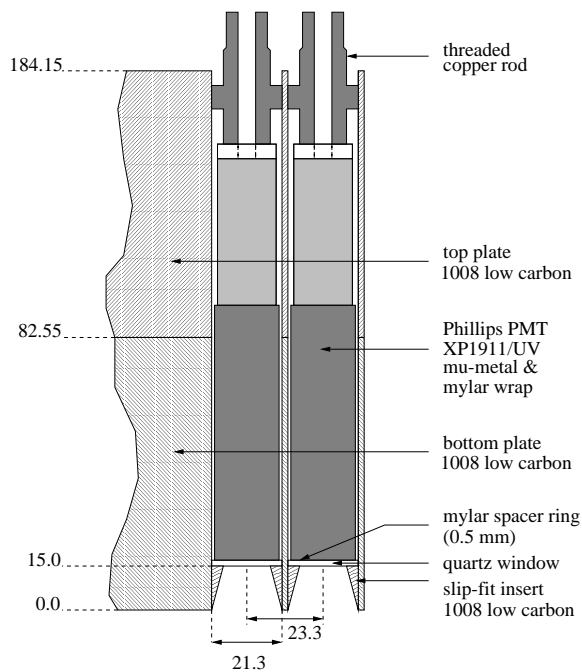}
\caption{Schematic photon detector design. All units are in mm.}
\label{fig:concept7}
\end{figure}

The photomultipliers were arranged in a hexagonal closed packed matrix 
of 147.4~cm by 62.8~cm, centered on the mirror focal point. 
This configuration was first used by the experiment 
E781-SELEX at FNAL~\cite{bib4}.  
The elementary cell of the array is a hexagon with the 
photomultiplier at the center. A light-gathering cone is used to increase the 
photon collection efficiency by reducing the dead space between the photocathodes. 
The distance between two adjacent cell centers 
is 23.3~mm, yielding an elementary cell 
surface of 426~mm$^2$. The 1934 PMTs
in each detector half are arranged in 73 columns of alternately 
26 and 27 PMTs each. A section of the soft steel matrix plate and
PMT packing design is shown in figure~\ref{fig:concept7}. 
The matrix plate provides the
gas seal for the C$_4$F$_{10}$ volume. The gas seal for individual PMT cavities
was provided by thin quartz windows which were glued to soft steel cone-shaped
inserts that in turn were glued into the inner face of the PMT matrix plate. 

Only 38\% of the area of the focal plane is covered by the PMT 
photocathodes. An aluminized plastic foil funnel 
was inserted into each soft steel entrance cone to
increase the coverage to 91\% of the photon detector surface and to improve the
reflectivity of the cone surface. 
\begin{figure}[h]
\vspace*{-0.5cm}
\includegraphics[width=3.2in]{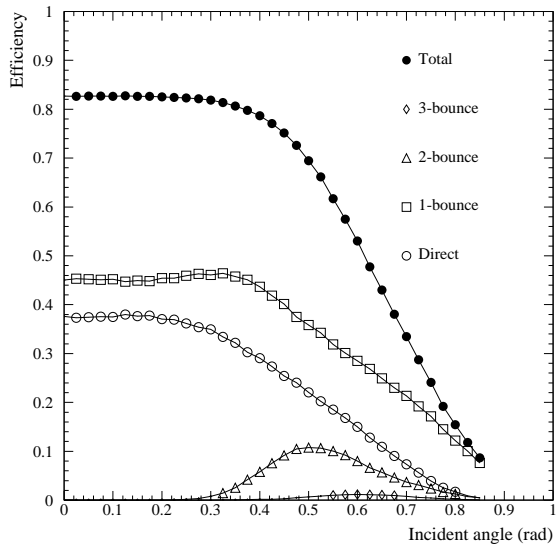}
\caption{Collection efficiency of the funnels versus incident
angle (solid points) from a Monte Carlo simulation. The other curves show the components
of the efficiency in terms of the number of ``bounces'', i.e. reflections
from the foil surface.}
\label{fig:funnel}
\end{figure}
The funnels extend beyond the steel
plate with an opening diameter of 23.3 mm, so minimizing the dead
space between the PMTs. They provide a high 
reflectivity above $\lambda$ = 200~nm. Their effect on the 
geometric collection efficiency
is shown in figure~\ref{fig:funnel} as a function of the angle between
the incoming photon and the normal to the detector plane (incident angle).
The influence of the funnels is found from the difference between the total
efficiency and the efficiency for "direct" detection.

The stray fields from the HERMES spectrometer magnet at the location of the 
photon detector matrix can reach up to 90~G perpendicular to the detector
plane, requiring careful magnetic shielding of the PMTs. 
Individual PMTs were wrapped in a 100$\mu$m thick $\mu$-metal
sheet and the matrix plates holding the PMTs were constructed from high 
permeability (C-1008) steel.
The soft steel matrix, combined with the
$\mu$-metal shielding and the soft steel inserts, reduces the magnetic fields
to negligible levels, which
guarantees that the PMT gains are not significantly affected.

\subsection{Detector Readout and Event Format}

The readout of the photon detector is performed by the commercial LeCroy
PCOS4 acquisition system, upgraded for the HERMES application.  Each detector
half is read out by a set of 8 symmetric backplane sections, each housing
15 or 16 PCOS4 cards. Each card in turn processes signals from 16 PMTs. 
This reduces the number of cables needed to 10 twisted pairs cables. 
Only digital information - when the pulse
exceeds the threshold of 0.1 photoelectrons -
is recorded.  The system is characterized by high input 
sensitivity (the threshold is 3000 electrons) 
and high amplification \linebreak ($-4.3$ $\mu$V per electron).
Because of the 650 W power dissipated per detector half, each RICH section
has a cooling system integrated into the RF-shield and electronic enclosure. 
A closed stream of cooling air (360 m$^3$/h) is guided over the PCOS4 electronics
on both sides of the PM-matrix and over the PMT matrix itself. The heat 
transferred to the air flow is removed by a water-cooled heat exchanger.

The data from both RICH detector halves are trea\-ted in the same manner as spectrometer
chamber data.  The PCOS4 system generates a data stream on an event-by-event
basis. The RICH data consists of a RICH hit table for each detector
half, where a PMT `hit' simply refers to a PMT that fired during the event time window.
A RICH mapping file is used to link the data channel numbers computed in the
decoding process to the PMT location in the detector matrix.  This information
can then be used to generate the spatial hit coordinates and the hit pattern
in the focal plane.

\section{Online Monitoring}

HERMES uses a client-server based system for the monitoring
of detector operating parameters~\cite{online,dad}. 
In the case of the RICH this includes high and low voltages, 
gas pressure and composition, and the temperatures at the 
photon detectors. The information is available in the form
of Tcl/Tk based displays~\cite{pink}.  

\begin{figure}[t]
\begin{center}
\includegraphics[width=2.8in]{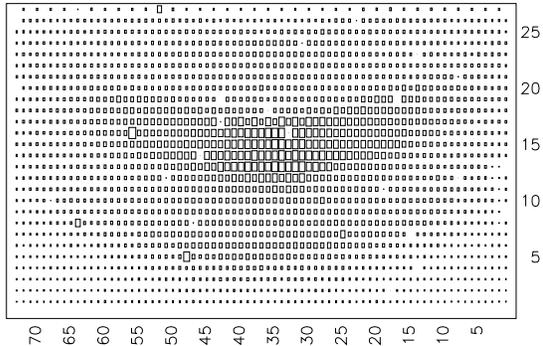}
\caption{Two-dimensional tubemap of relative yield 
for the bottom RICH detector.}
\label{fig:tubemap}
\end{center}
\end{figure}

In analogy to wiremaps for the tracking chambers, ``tubemaps''
are used to monitor the performance of the PMTs, the high voltage
and the electronics. An example of the standard two-dimensional
overview is shown in figure~\ref{fig:tubemap}.
The increased hit density in the center is due to the gas rings.
This plot shows that more than 99\% of the tubes are functional.
More detailed maps are available for diagnosing operational problems.

\begin{figure}[h]
\begin{center}
\includegraphics[width=2.8in]{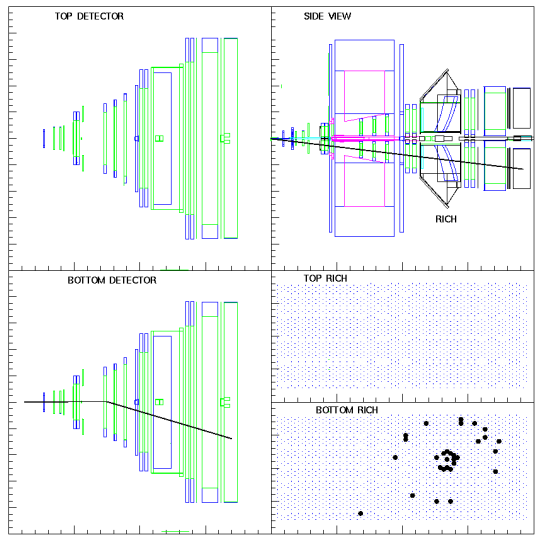}
\caption{Online display of a single event.}
\label{fig:eve1}
\end{center}
\end{figure}

The RICH data have also been integrated into the online HERMES event 
display system to allow the online analysis of events. Figure~\ref{fig:eve1} 
shows the standard view for the case of a clean single electron event.
The event display combines top and side views of the spectrometer with
views of the top and bottom RICH photon detectors.
It represents the grid of 1934 PMTs in the form of dots. All
PMTs that fire are marked by filled circles. In general particles with momenta 
above both the aerogel and ${\mathrm C_{4}F_{10}}$ thresholds will
generate roughly concentric rings: a small inner ring resulting from
the photons generated in the gas and a larger outer ring formed 
by the aerogel photons.

\begin{figure}[h]
\begin{center}
\includegraphics[width=2.8in]{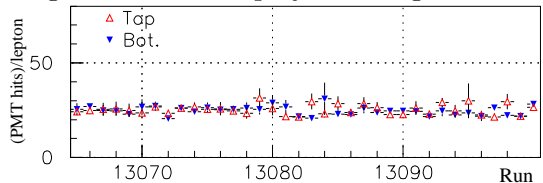}
\caption{
Average number of hit PMTs per single lepton
event plotted versus run number.
}
\label{fig:daql}
\end{center}
\end{figure}

Electrons and positrons that are detected in the RICH 
always have momenta above both radiator thresholds. 
They should therefore always
have both rings present, which makes them ideal for the online
monitoring of the data quality.
The average number of PMT hits for single lepton events in a run is 
used as the monitoring parameter;
it is usually about 20. 
Figure~\ref{fig:daql} shows this parameter in a typical online 
data quality plot.

\section{Reconstruction and Particle Identification}

\subsection{Principles of the Reconstruction}
\label{sec:reconstruction}

Since the sensitive face of the flat photon detector does not conform to the
true mirror focal surface, the detected `rings' are not circles, but in general
asymmetrically distorted ellipses. Hence a ring fitting procedure is not
feasible for an accurate reconstruction of the \v{C}erenkov angles. 
The raw data that provide the input for any reconstruction procedure are
the list of PMT hits per event and the parameters of the particle tracks 
determined by the tracking detectors~\cite{spectrometer}. 

In the case of the aerogel rings on average only about 10\% of the PMTs 
that lie on the aerogel ring will be hit. Hence multiple hits are infrequent and
each PMT hit represents one photoelectron. The average number of hits on
the aerogel ring of an ultra-relativistic ($\beta\approx 1$) particle that
suffers no acceptance effects (aerogel tiles, mirror) is about 10.
The sparsity of the hits on the aerogel ring makes the detection of this ring
susceptible to background, noise and acceptance problems.
The situation for gas rings is very different.
The average multiplicity of photoelectrons per hit for the gas rings is 
about 2 (based on MC simulations). This might lead to the assumption 
that all PMTs that lie on the gas ring
will be hit. However, due to Poisson-statistics this is not correct. 
On average about 23 PMTs are touched by
the gas ring of an ultra-relativistic particle, but 
only 12 of those PMTs actually fire.
Most of them are hit by several photo-electrons.

\begin{figure}[t]
\begin{center}
\includegraphics[width=3.0in]{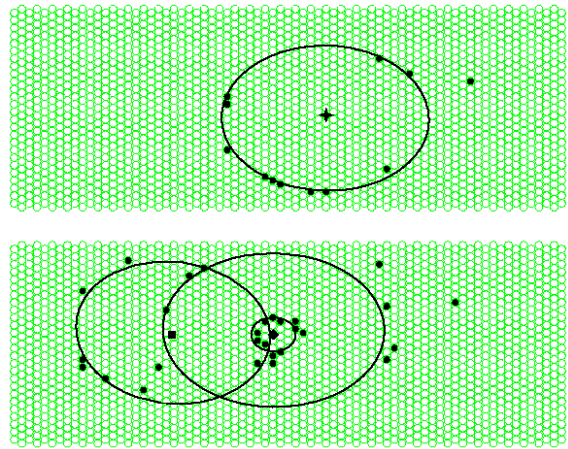}
\end{center}
\caption{HERMES RICH event display for an event with a 14.6 GeV electron (right)
and a 1.5 GeV $\pi^-$ (left) in the lower half and a 5.5 GeV K$^+$ in the upper half. 
See text for detailed description.}
\label{fig:event}
\end{figure}

Figure~\ref{fig:event} shows the offline RICH event display for a three-track
event that illustrates several typical features of HERMES RICH events. 
The event shown has a 14.6 GeV electron\footnote{Electrons and positrons are identified by
a combination of calorimeter, preshower and TRD with an average
efficiency of 99\% and a hadron contamination below
1\%~\cite{spectrometer}.} and a 1.5 GeV pion in the lower half,
and a 5.5 GeV kaon in the upper half of the detector. The solid black points
mark the PMT hits, while the markers in the ring centers indicate the
virtual track hit points, i.e. the points where the particle tracks would 
intersect the photon
detector if they were imaged by the mirror. The solid lines are spline fits
to a few simulated photon hits. They indicate where, based on the track
parameters and the particle type, hits 
could be expected for this event.
The electron track is easily identifiable as the only one with a gas ring;
a comparison to figure~\ref{cerangles} shows that only for the electron
a gas ring is expected. The momentum of the pion is below the pion gas threshold 
and thus it only exhibits an aerogel ring. However, the particle clearly must be 
a pion, 
because the particle momentum is below the aerogel threshold for kaons. The kaon in
the top detector has a well defined aerogel ring, but no gas ring - as is
expected for a kaon at 5.5 GeV. Any pion at this momentum would
certainly have produced a gas ring, while the aerogel ring for a
proton would have a much smaller radius.

The task of the pattern recognition algorithm is in principle to
associate the various hits in the photon detector of the RICH with a ring assigned
to a certain track. In practice this detailed assignment is not needed
and only the information relevant to the track identification is extracted.
The relatively low track multiplicity in the HERMES
experiment simplifies this task considerably.

\subsection{Inverse Ray Tracing}

The analysis of the hit patterns is intrinsically complex since the
non-linearities of the imaging system distort the simple ring structure 
of the emitted light. The influence of the imaging system can be removed
by {\em inverse ray tracing} (IRT)~\cite{forty,bib2}. 
In this method the \v{C}erenkov angle
corresponding to each PMT hit is reconstructed from the track parameters
and the position of the PMT.

The inverse ray tracing problem can be formulated as follows:
given a track and a hit in the RICH photon detector plane, 
at which angle was the photon emitted? Assume that 
the emission point can be estimated. This assumption will be
discussed below.
\begin{figure}[hbt]
\begin{center}
\includegraphics[width=2.8in]{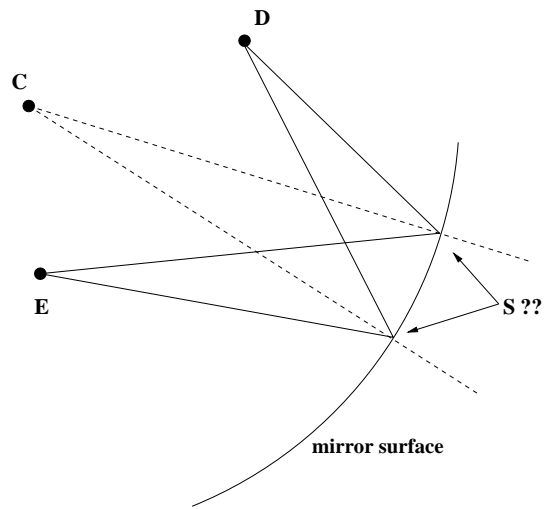}
\end{center}
\caption{The problem of inverse ray tracing.}
\label{fig1}
\end{figure}

The geometrical problem can be formulated as follows using the
terminology of figure~\ref{fig1}. Given point
E, the likely emission point, point D, the detection point 
and C, the center of the spherical mirror the photon scatters from, find
the point S on the surface of the mirror where the photon scattered. 
The properties of point S are, in vector notation:
\begin{enumerate}
\item $\vec{CS}$ is coplanar with $\vec{CE}$ and $\vec{CD}$.
\item $|\vec{CS}|$ = R
\item $ \sphericalangle ( \vec{ES}, \vec{CS} ) = 
\sphericalangle ( \vec{CS}, \vec{DS} )$
\end{enumerate}

\begin{figure}[t]
\begin{center}
\includegraphics[width=2.in]{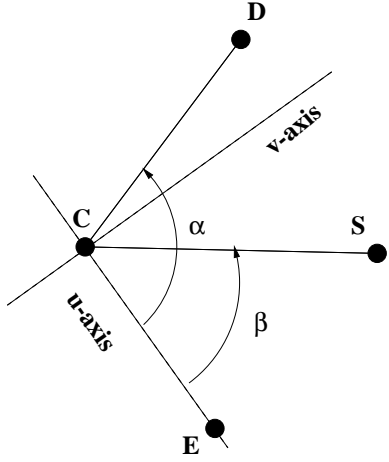}
\end{center}
\caption{Definition of the base and the angles}
\label{fig2}
\end{figure}

For the mathematical formulation of the problem it is easier to switch
to an Euclidean base with C as the origin. 
The u axis is defined along $\vec{CE}$. The v axis is
coplanar with $\vec{CE}$ and $\vec{CD}$, and oriented 
such that $\hat{v}\cdot\vec{CD} > 0$. The usual {\em caret} 
notation is used to indicate a unit vector.
The third axis (w) is defined such that the base defines a right-handed coordinate
system. The components of the vectors are then:
\begin{alignat}{3}
\vec{CE}=(& a ,& 0 ,& 0)\nonumber \\
\vec{CD}=(& d\cos\alpha,& d\sin\alpha ,& 0) \nonumber \\
\vec{CS}=(& R\cos\beta ,& R\sin\beta,& 0)
\end{alignat}
Clearly $|\vec{CE}| = a$ and $|\vec{CD}| = d$.
The angles are defined in figure~\ref{fig2}. 

The coplanarity requirement has been satisfied by the choice of the axes.
The para\-meterization of S ensures that it will be located on the mirror surface. 
The remaining parameter $\beta$ now must be determined from the 
equality of the incident and reflected angles. This leads to two
(redundant) equations for the unit vectors:
\begin{eqnarray}
\hat{SC} \cdot \hat{SE}  =& \hat{SD} \cdot \hat{SC} \\
\hat{SC} \times \hat{SE} =& \hat{SD} \times \hat{SC} 
\end{eqnarray}
The second equation is a vector equation, but since all three vectors
are coplanar, the only parameter is the length of the vector product. 

The first equation reduces to:
{\footnotesize
\begin{equation}
 \frac{|\vec{SD}|}{|\vec{SE}|} = 
\frac{d\cdot(\cos\alpha\cos\beta+\sin\alpha\sin\beta)-R}
{a\cos\beta- R} 
\end{equation}
}
The vector equation represents the equality of the sines of the angles,
and has only a non-zero component along the w-axis. This component reduces
to:
{\footnotesize
\begin{equation}
\frac{|\vec{SD}|}{|\vec{SE}|} = 
\frac{d(-\cos\alpha\sin\beta+\sin\alpha\cos\beta)}{a\sin\beta }
\end{equation}
}
Combining the two equations yields:
{\footnotesize
\begin{multline}
ad\cdot\sin(\alpha - 2\beta)+\\
R\cdot(a\sin\beta- d\sin(\alpha -\beta))=0
\end{multline}
}
The solution to the equation can now be found from Newton-Raphson iterations. 
\newline
Since the result will be the angle $\beta$, no further work is necessary. Using 
half of the angle $\alpha$ as the starting point, convergence is usually
achieved within three iterations. 
The reflection point on the mirror can then be found as follows:
{\footnotesize
\begin{multline}
\vec{S} = \vec{C}+
( \frac{R\cos\beta}{a} - \frac{R\sin\beta\cdot\cos\alpha}{a \cdot \sin\alpha}) 
\cdot\vec{CE}\\
+\frac{R\sin\beta}{d\sin\alpha}\cdot\vec{CD}
\end{multline}}

The inverse ray tracing method reconstructs the emission angle for each
hit, provided the emission vertex is known. 
For each track it is assumed that the hit could be 
coming from the aerogel or the gas, and the emission vertex is estimated 
\begin{figure}[b]
\begin{center}
\includegraphics[width=3.0in]{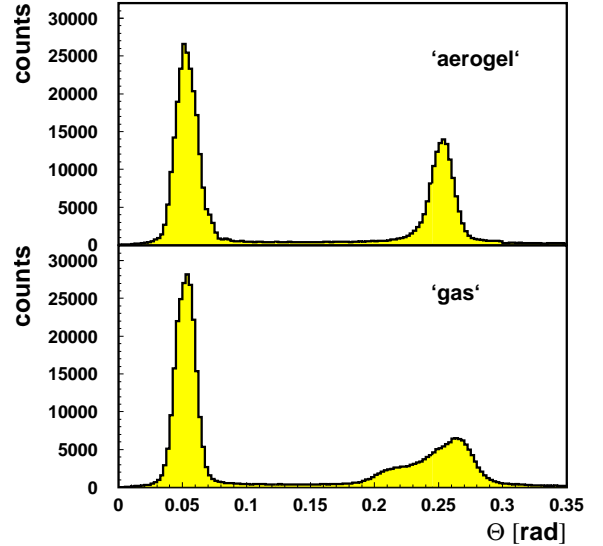}
\end{center}
\caption{Reconstructed angle spectra for the assumption that the photon
was emitted in the aerogel (top) or in the gas (bottom). The data are from
single track, low background electron events with $p>5$ GeV.}
\label{fig:rec_angle}
\end{figure}
accordingly. For each radiator hypothesis, the emission angle is
then reconstructed. 
This results in a spectrum of reconstructed angles, as shown in
figure~\ref{fig:rec_angle}, which is clearly dependent on the radiator 
hypothesis. Due to the very different index of refraction, 
it is rarely a problem to distinguish which radiator hypothesis 
is correct. The remaining uncertainty in the emission vertex due to
the finite length of the radiator is unavoidable and results in
the contribution $\Delta\theta_{em}$ to the resolution in table
\ref{table:delta}.

Another difference in the treatment of reconstructed aerogel and gas angles
lies in the refraction of aerogel photons at the aerogel-gas
boundaries. This is approxima\-tely corrected for under the 
assumption of a flat aerogel-gas boundary and coplanar refraction.

\subsection{Likelihood Analysis}

The selection of a most probable particle type is made by
calculation of the likelihood that each particle type 
would generate the measured IRT spectrum.
Due to aerogel edge effects and the finite mirror size, 
the yields of aerogel and gas hits vary strongly over the 
spectrometer acceptance. 
As a result the particle identification is largely based on the 
reconstruc\-ted average angles, which are less susceptible to these effects.
The theoretically expected angles $\theta_{th}$
are calculated for each particle type hypothesis $i$ 
from the track momenta. For each particle type hypothesis 
a window is imposed on the spectrum of reconstructed angles around 
this expected value. The width of this window ($\sigma_w$) 
is set to 4 times the single photon resolution $\sigma_\theta$.
However, it is in principle a free parameter that can be adjusted to
optimize the performance of the particle identification algorithm.
The average angle is then determined from the reconstructed angles in 
the window $[\theta_{th}-\sigma_w;\theta_{th}+\sigma_w]$.

The distribution of measured average angles for a given particle type
and momentum can be normalized to form a conditional probability.
For known relative particle fluxes a particle type probability can then be
calculated. If the relative particle fluxes are not known, the conditional
probability can itself be used as a likelihood. Furthermore, if the
average angle resolution is assumed to be independent of the particle
type, the overall normalization of the conditional probability
distribution does not matter. It may be chosen so that the maximum
value of the likelihood is 1. With a Gaussian shape of the average
angle distribution the likelihood is then calculated as
\begin{equation}
L(\langle \theta \rangle)=exp\left[ -\frac{(\theta_{th}-\langle \theta \rangle)^2}
{2\sigma_{\langle \theta \rangle}^2}\right]
\end{equation}
where $\sigma_{\langle \theta \rangle}$ is the average angle
resolution, which is calculated from the single photon
resolution as
\begin{equation}
\sigma_{\langle \theta \rangle}=\frac{\sigma_{\theta}}{\sqrt{N}}
\end{equation}
The likelihoods for the two radiators, aerogel and gas, are combined
to an overall likelihood by multiplication.
For this paper, the particle is assigned the type with the highest likelihood. 
Obviously, this requirement
will allow more contamination than is acceptable for most physics analyses. A 
detailed discussion of the ``quality'' of the particle identification
and the resulting trade\-off between contamination and efficiency
will be the subject of a subsequent paper.

\section{Mirror Alignment}

\label{alignment_section}
The alignment of the mirror array within the detector volume was
determined from the data. It consisted of three steps: global
alignment, individual mirror alignment and local alignment.
 
\begin{figure*}[htb]
\begin{center}
\epsfysize=5.5in \leavevmode \epsfbox{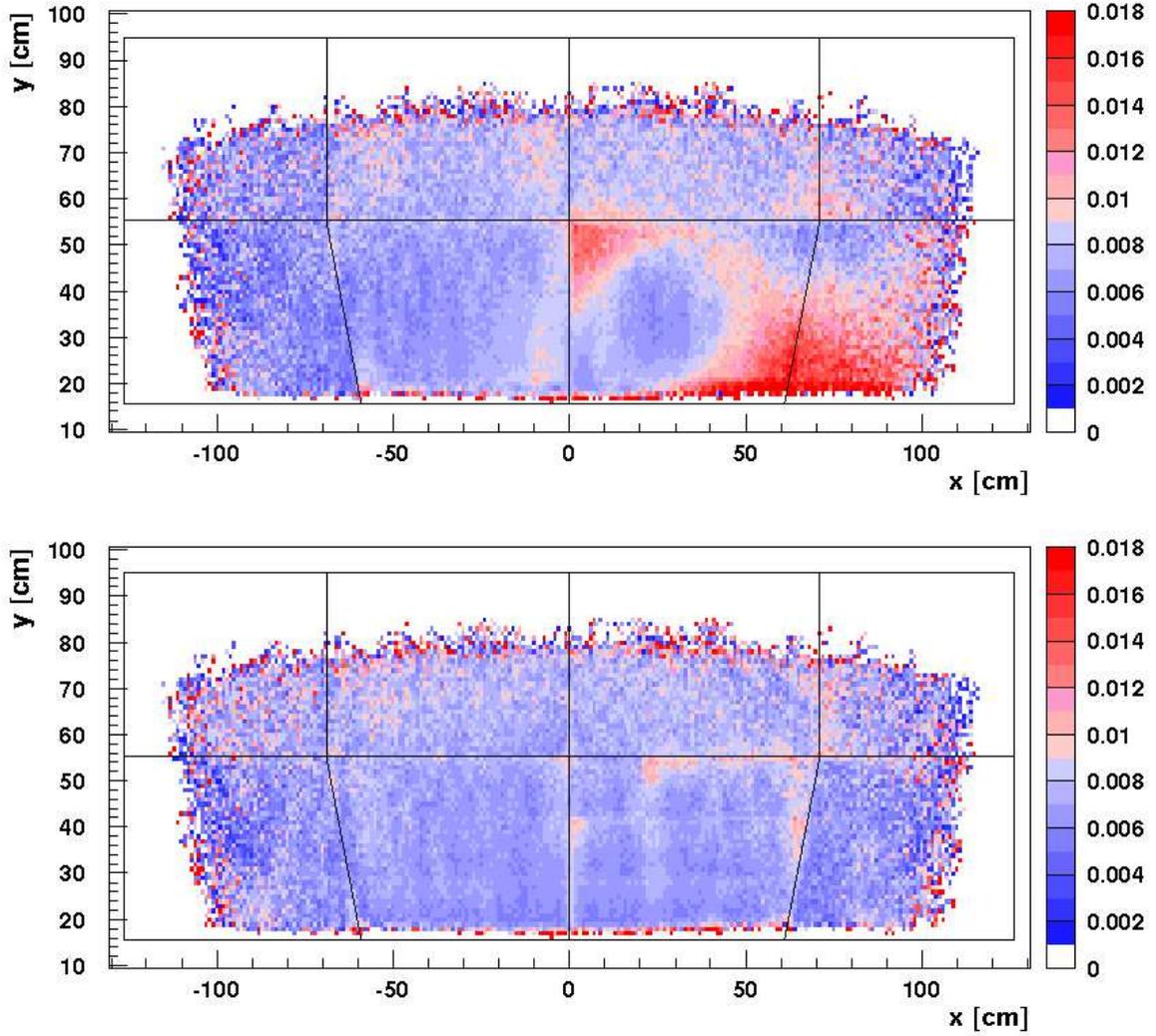}
\caption{Focusing quality $| \theta_i-\langle \theta \rangle|$ for pion gas rings
(4.5-5.5 GeV) before and after the individual and local mirror alignment. 
The lines indicate the boundaries
of the individual mirrors.}
\label{focus}
\end{center}
\end{figure*} 
 
For global mirror alignment the mirror array was treated as one unit.
A three-dimen\-sional scan of the assumed position of the mirror
center was performed (in software) 
and the influence on the reconstructed aerogel and
gas rings was studied using a set of high energy single electron tracks. The
performance was characterized using 24 parameters, including the average angles,
their spread, the single photon resolution, the yields and the azimuthal 
uniformity of the gas rings. A careful study 
resulted in a small range of acceptable values for the position of the mirror
center and a shallow optimum was selected. 

However, both mirror arrays consist of eight individual mirrors each, whose
alignment relative to one another is not necessarily perfect.
In addition, the construction of the mirror allows for 
local variations of the radius
of curvature of the surface, especially around the stress points
introduced to compensate for the flexibility of the individual mirror segments
(positioned at the edges of the mirrors). These variations in the mirror radius 
introduce local variations in the reconstructed angle. This translates into a larger
spread on the reconstructed angle, and (if not corrected) 
a different average angle for some parts of the mirror. 
In certain momentum ranges this can lead to a 
misidentification of the particle. 

For this reason in the next two steps of the alignment determination
the individual mirrors are aligned relative to one another and then 
each mirror is further subdivided and the pieces are individually adjusted.
Steps two and three are based on two geometric properties
of the mirrors: the position of a  spherical mirror with \linebreak known radius 
is unambiguously determined by the position of the mirror center. 
Furthermore, if it is known how a single photon is reflected by the
mirror, the mirror center can be determined for a given radius.

While the actual track of a photon in the detector cannot be determined,
it is possible to determine approximately how a charged particle track would
reflect in the mirror. This is done by measuring the center of the gas
ring created by the track. This center can be related
to the point where a reflected track would hit the photon detector
by comparing it to an MC simulation. 
For a single track the reconstruction of the individual mirror centers
is still under\-de\-ter\-mined. However, it is possible to use a large number 
of tracks with gas rings that were reflected on the same individual 
mirror and then calculate the optimal mirror center to account for 
their behavior. This optimum is not unique and depends somewhat on the
starting point. 

For the third step the individual mirrors were artificially divided
into $3\times 3$ parts that were treated as if they were individual mirrors.
The limiting factor here is the size of the gas rings. 
This `local alignment' allows for a better approximation of the real 
mirror surfaces.
The result is illustrated for the top mirror array in figure
\ref{focus}. The plotted quantity is the focusing quality \newline
$| \theta_i-\langle \theta \rangle|$, where
$\theta_i$ is the 
reconstructed \v{C}erenkov angle for a single PMT hit and $\langle \theta
\rangle$ is the average angle for the corresponding track. 
This quantity approximately corresponds to a local derivative 
of the mirror surface.
Individual mirror and local alignment are based on a large sample of pions with
momenta between 4.5 and 5.5 GeV. They were selected to cover as much as 
possible of the mirror surface while having well defined gas rings.

\section{Detector Performance}
\enlargethispage{\baselineskip}
\label{performance}
The detector performance ultimately is measured by how well the various
particle types are identified. 
For a given particle identification algorithm 
this performance is determined by the number of detected gas
and aerogel photons, as well as the single photon resolution. Complications
arise from overlapping rings and background hits.
\begin{figure}[t]
\epsfig{file=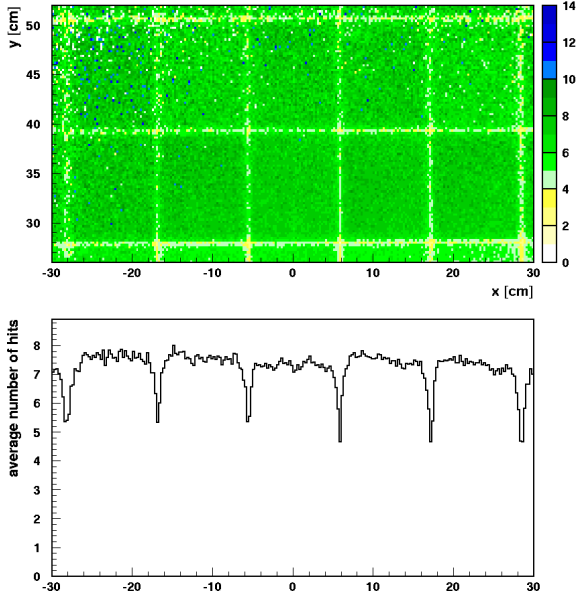 ,width=3in}
\caption{Dependence of electron aerogel yield on track position at 
the aerogel radiator.}
\label{aero2d}
\end{figure}

\subsection{Yields}
The number of aerogel photons detected strongly depends on 
how much the track is affected by acceptance effects from the finite size
of the mirror array and the tile structure of the aerogel radiator.
Figure~\ref{aero2d} illustrates the influence of the tile structure.
The bottom plot is a projection of of a horizontal band of tile in the
top plot between y=30 cm and 36 cm,  
showing the edges of the aerogel tile stacks. 
The slightly lower average yield in the top left tile 
arises from a lower incident electron flux.
With the given size of 
the aerogel tiles about 45\% of all tracks are affected by the tile
structure. It would be desirable to develop larger aerogel tiles of
the same optical quality. For example, in the case of 20$\times$30~cm$^2$  
tiles only about 20\% of the tracks would be affected.

\begin{figure}[t]\centering
\epsfig{file=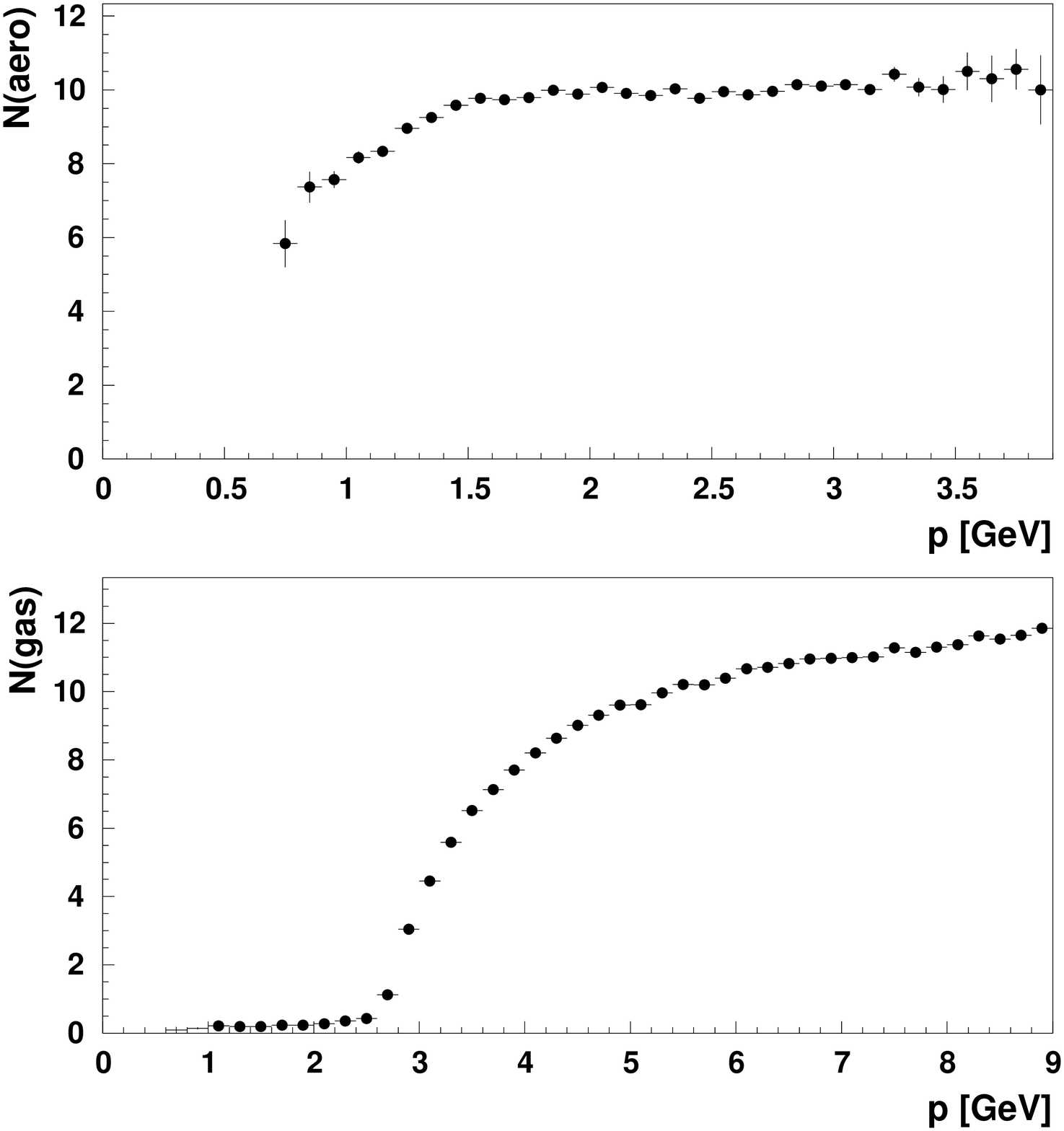 ,width=2.8in}
\caption{Number of fired aerogel (top) and gas PMTs (bottom) versus particle momentum for 
`ideal' pion tracks.}
\label{npmt}
\end{figure}

Figure~\ref{npmt} shows the number of aerogel
PMT hits ${\mathrm N(aero)}$ and gas PMT hits 
${\mathrm N(gas)}$ versus particle momentum 
for `ideal' pion \linebreak tracks that do not suffer from acceptance or overlap effects. 
The two plots show the rise towards the asymptotic values of 10 aerogel hits 
and 12 gas hits. As a result of the acceptance effects, the asymptotic number of
aerogel hits for all tracks is reduced to about 8.

\subsection{Background}

An inspection of all events from a single
run (1861 events) showed that roughly one third of all particles
that produce clear rings in the detector do not
have a full reconstructed track that is associated with them. 
These tracks are low momentum particles that do not pass through the entire 
spectrometer and particles produced in flight within the spectrometer. 
Examples of the latter are delta electrons and electron-posi\-tron
pairs. These `trackless rings' act as background for the rings that are
associated with tracked particles. 
In events with more than one track
in one detector half the rings of the tracked particles of course are
`background' for one another in the same way.

In addition to the trackless rings, there are several sources
of background photons that do not necessarily result in ring structures.
These include Rayleigh scattered photons, \v{C}erenkov photons produced 
in the lucite window, proton beam correlated back-{\linebreak}ground showers that 
hit the PMT matrix directly and scintillation in the gas.
The electronic and PMT noise in the detector is a very small effect,
amounting to only about 1 fired PMT every 5 events.

\subsection{Average Angles and Resolution}
\begin{figure}[b]\centering
\epsfig{file=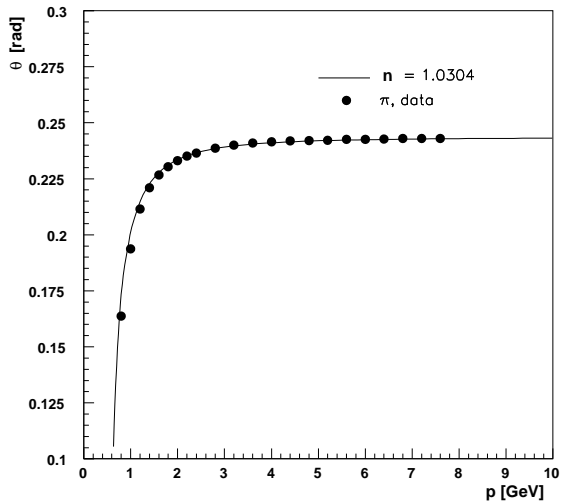 ,width=2.9in}
\caption{Reconstructed average aerogel angle versus particle momentum for pions. 
The solid line represents a fit with n=1.0304.}
\label{nbot}
\end{figure}

\begin{figure}[hb]\centering
\epsfig{file=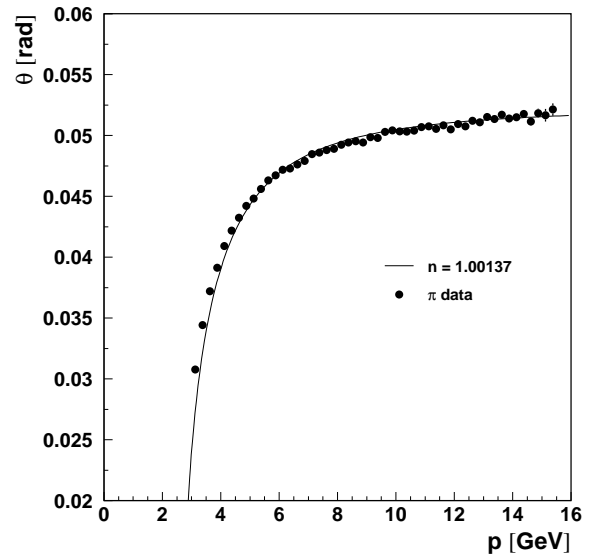 ,width=3.in}
\caption{Reconstructed average gas angle versus particle momentum for pions. 
The solid line represents a theoretical curve with n=1.00137.}
\label{gasangles}
\end{figure}

As the likelihood analysis is based on the average angles, it is of particular
importance that the angles are correctly reconstructed and that their 
momentum dependence is understood. Figure~\ref{nbot} shows the reconstructed
average aerogel angle for pions. The data were fit with a theoretical curve
with the aerogel index of refraction as the only free parameter. The resulting
curve and index of refraction (n=1.0304) 
are in excellent agreement with the theoretical expectation and the
optically measured index of refraction.
Figure~\ref{gasangles} shows the reconstructed gas angles for pions
together with the theoretical curve for an index of refraction of
1.00137. The systematic difference between data and theoretical curve
at low momenta is due to the finite size of the PMTs. 

\begin{figure}[h]\centering
\vspace*{-0.5cm}
\epsfig{file=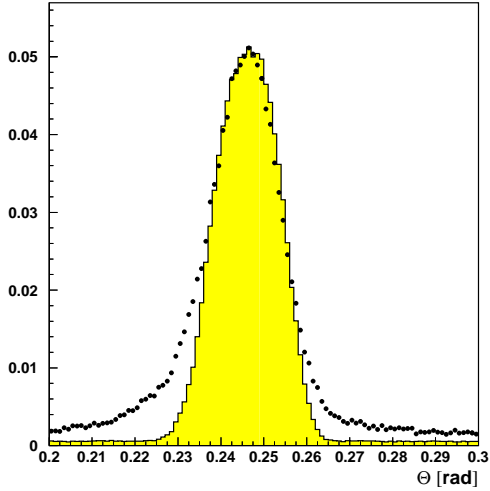 ,width=2.9in}
\caption{Normalized distributions of reconstructed aerogel angles for single, low background
electrons ($p>5$ GeV). Histogram: MC, points: data.}
\label{aeromcdata}
\end{figure}

\begin{figure}[h]\centering
\vspace*{-0.5cm}
\epsfig{file=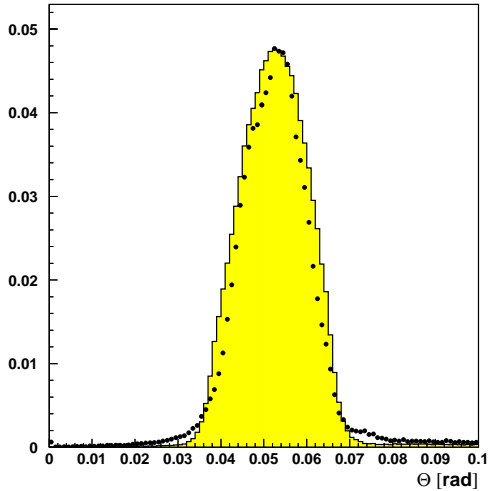 ,width=2.9in}
\caption{Normalized distributions of reconstructed gas angles for single, low background
electrons ($p>5$ GeV). Histogram: MC, points: data.}
\label{gasmcdata}
\end{figure}

\enlargethispage{\baselineskip}
The MC simulation of the RICH detector plays an important
role in the understanding of efficiency and contamination values for
the RICH particle identification. Hence, it is necessary that the
detector response is adequately described by the simulation.
Figures \ref{aeromcdata} and \ref{gasmcdata} show the comparison of
reconstructed aerogel and gas angles for ultra-relativistic electrons \linebreak
($p>5$~GeV).

For the comparison the data and the MC results
were normalized to the same maximum value. 

Figure~\ref{aeromcdata} shows the excellent agreement
in the central part of the distribution. 
The wider tails of the experimental distribution are due to
several effects that will be discussed below. The flat 
background past the tails is another important difference 
between the MC and the data. 
The difference in the aerogel spectra is due to the presence of additional
contributions to the resolution, which are listed in 
table~\ref{table:delta2}. A significant effect comes from
the aerogel tiles, which have sharply curved surfaces 
near the edge. This refracts the photon in an unpredictable way, which
gives rise to the $\Delta \theta_{tile}$ contribution. It also
contributes to the background.
$\Delta {\theta}_{tile}$ was estimated from comparing
the resolution for tracks that are not affected by the tile edges to 
those that are. 
$\Delta {\theta}_{nvar}$ accounts for variations in the 
index of refraction between the aerogel tiles. Its value is
determined from the measured indices of refraction. 
There is also a `forward scattering' contribution $\Delta{\theta}_{fw}$
\cite{forward}, which must be taken into account, but which, unlike
the contribution $\Delta {\theta}_{chr}$, is not included in the
Monte Carlo simulation.

\begin{table}[b]\centering
\begin{tabular}{|l||c|c|}
\hline
~ & aerogel & ${\mathrm C_4F_{10}}$ \\
\hline
$\Delta {\theta}_{tile}$ & $\sim$3.0 mrad & - \\       
$\Delta {\theta}_{nvar}$ & $\sim$1.1 mrad & - \\       
$\Delta {\theta}_{fw}$ & $\sim$0.9 mrad & - \\       
$\Delta {\theta}_{press}$ & - & $\sim$1.0 mrad \\       
\hline
$\Delta {\theta}_{exp}$ & 7.6 mrad & 7.5 mrad \\ 
\hline  
$\Delta {\theta}_{mirr}$ & $\sim$2.3 mrad & $\sim$2.2 mrad \\
\hline
\end{tabular}
\vspace*{0.2cm}
\caption{Additional contributions to the single photon resolution in
aerogel and ${\mathrm C_4F_{10}}$. The contribution to the resolution
related to the mirrors ($\Delta {\theta}_{mirr}$) has been obtained
by subtracting the other relevant contributions from $\Delta {\theta}_{exp}$.}
\label{table:delta2}
\end{table}

For the reconstructed gas angles in figure \ref{gasmcdata},
the width and shape of the 
distributions is basically in good agreement. The MC distribution
is slightly wider and less `Gaussian' than the experimental data.
As this is the distribution of the individual reconstructed
angles (with one entry per fired PMT), this is likely an effect
of the MC overestimating the photoelectron yield for the gas. 
Due to the non-linear relationship between the number of
photoelectrons and the number of fired PMTs the additional
photoelectrons add relatively more to the sides than to the center
of the distribution.

Pressure and temperature fluctuations \linebreak change the density of the 
gas, and therefore its index of refraction. This contribution,
$\Delta {\theta}_{press}$, was determined from the observed maximum
pressure fluctuations. This error can be eliminated by taking the measured
pressures into account in the reconstruction.

The experimental value of the total angular resolution for single
photons, $\Delta {\theta}_{exp}$, 
was determined for high energy single lepton tracks that are not 
affected by any tile edge or mirror acceptance effects 
(table~\ref{table:delta2}). 
The data were taken from a period of stable atmospheric pressure and
events with a large background were excluded. The reconstruc\-ted
angle spectra were fitted with a Gaussian plus a linear background.
As a result of the selection of these `ideal' tracks the aerogel
tile effects ($\Delta{\theta}_{tile}$) do not contribute to 
$\Delta {\theta}_{exp}$ for aerogel and $\Delta {\theta}_{press}$
does not contribute to $\Delta {\theta}_{exp}$ for gas.
The experimental gas resolution therefore is a combination of 
the effects that are collected in $\Delta {\theta}_{MC}$ 
(cf. table~\ref{table:delta}) and the hitherto unknown contribution
of the mirror imperfections $\Delta {\theta}_{mirr}$.
This contribution in particular refers to the non-sphericity of the mirror array and 
the diffuse reflection component of its surface.
Under the assumption that the quadratic sum of $\Delta {\theta}_{MC}$ and  
$\Delta {\theta}_{mirr}$ yields the experimental value 
$\Delta {\theta}_{exp}$ for gas, the mirror contribution is calculated to be
about 2.2~mrad. In the case of the aerogel also $\Delta {\theta}_{nvar}$
and $\Delta {\theta}_{fw}$ must be included. 
The assumption that $\Delta {\theta}_{exp}$ for aerogel is the quadratic
sum of these contributions with $\Delta {\theta}_{MC}$
(cf. table~\ref{table:delta}) leads to a value of 2.3~mrad
for $\Delta {\theta}_{mirr}$, in good agreement with the value
extracted from the gas angles.

A comparison of tables \ref{table:delta} and \ref{table:delta2}
shows that the dominant contribution to the resolution comes
from the pixel size ($\Delta{\theta}_{pix}$).

\subsection{Efficiencies}
Monte Carlo predictions for the efficiencies and purities
of the hadron identification using the IRT likelihood algorithm
are given in table~\ref{table:efficiencies}. The MC
simulation of the RICH detector used for this prediction 
contains a detailed simulation of the aerogel radiator geometry
as well as experimental background data extracted from 
trackless events. The data are from 1.84 million MC events
using a deep inelastic scattering generator. The particles are
identified as the particle type with the highest IRT likelihood.
As mentioned before, stronger constraints on 
the particle likelihoods will improve the purity of the sample,
albeit at the expense of the efficiency. \\

{\small 
\begin{table}[ht]\centering
\begin{tabular}{|c||r|r|r||c|}
\hline
id. as & $\pi$ & K & p & purity \\
\hline
\hline
$\pi$  & 619390 &  8050 &  23008 & $0.95$ \\
K      &  17282 & 58390 &  16757 & $0.63$ \\
p      &   7461 &  5612 &  80113 & $0.86$ \\
no id. &  10802 &  6647 &  29634 &   \\
\hline
\hline
$\varepsilon$ & $0.95$    & $0.74$     & $0.54$ & \\
\hline
\end{tabular}

\caption{Typical efficiencies and purities for the RICH hadron identification
based on the IRT likelihood analysis of 1.84 million DIS MC events,
${\mathrm p>2}$ GeV.}
\label{table:efficiencies}
\end{table}
}

\begin{table}[h]\centering
\begin{tabular}{|l||l|}
\hline
Decay  & Efficiency \\
\hline
\hline
$\rho^0 \rightarrow \pi^+\pi^-$ & $\varepsilon_{\pi}=0.915\pm0.024$ \\
\hline
$K_s \rightarrow \pi^+\pi^-$ & $\varepsilon_{\pi}=0.900\pm0.005$ \\
\hline
$\phi \rightarrow K^+K^-$ & $\varepsilon_{K}=0.750\pm0.007$ \\
\hline
$\Lambda \rightarrow p\pi^-$ & $\varepsilon_{p}=0.726\pm0.010$ \\
                             & $\varepsilon_{\pi}=0.890\pm0.011$ \\ 
\hline
\end{tabular}
\caption{Typical momentum integrated efficiencies  
 determined from decaying particles based on the IRT likelihood analysis.} 
\label{table:decays}
\end{table}
It is not possible to use another detector to create clean hadron
samples to study the detector performance. However, it is possible to
use decaying particles for the same purpose. Samples of
$\rho$, $\phi$ and $K_s$ mesons as well as $\Lambda$ hyperons were used 
to determine the identification efficiencies for pions,
kaons and protons. The momentum integrated results (p$>$2~GeV) 
are shown in table~\ref{table:decays}. 
The MC values in table~\ref{table:efficiencies} and the
experimental values in table~\ref{table:decays} cannot be compared
directly, because the efficiencies depend strongly on the particle
momentum as well as on the event topology.

\begin{figure}[ht]\centering
\hspace*{-0.3cm}
\epsfig{file=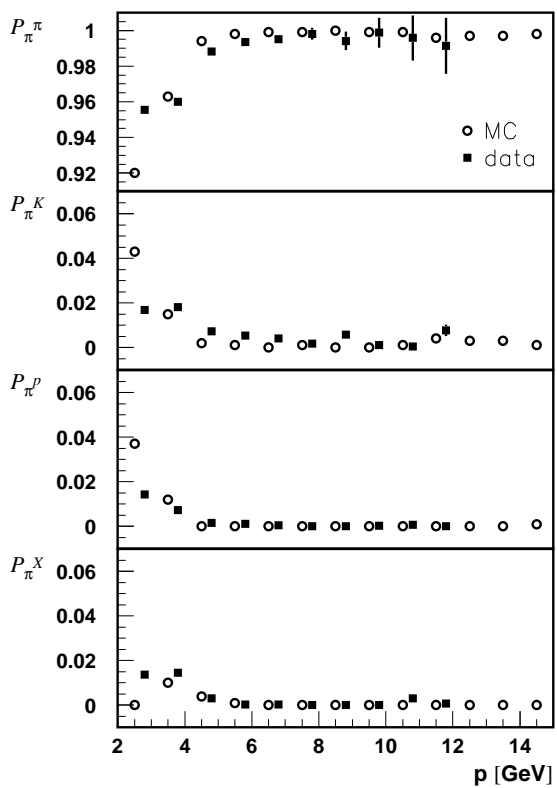 ,width=3.1in}
\caption{Identification probabilities $P_{\pi}^i$ that a pion
is identified as pion, kaon, proton or not identified ($X$)
for single tracks per detector half. DIS MC simulation (circles) in
comparison with experimental data from $\rho^0$-decays (solid
squares); both results are based on the IRT likelihood analysis.}
\label{ppi}
\end{figure}
This dependence on the event topology can be excluded by comparing
subsamples with only one track per detector half. 
Figure~\ref{ppi} compares MC and experimental data for the 
case of single track pions. The four plots show the identification
probabilities $P_{\pi}^i$ that a pion is identified as pion, kaon, 
proton or not identified ($X$). The pion identification
efficiency $P_{\pi}^{\pi}$ for single tracks is above 92\% even at 
low momenta, and above 98\% for momenta larger than 4 GeV. 
The MC data are taken from the sample also used for 
table~\ref{table:efficiencies}, while the experimental data are from 
$\rho^0$-decays. MC and experimental results agree well,with the
possible exception of the lowest moment bin. 

The comparison of MC and experimental data becomes more
involved for event topologies with overlapping rings and
as the opening angle for $\phi$-decays in HERMES is small
there are basically no experimental data  
with one single kaon per detector half.
A detailed evaluation of the detector performance that takes 
the event topology and momentum dependence into account 
in detail will be included in a forthcoming paper.

\section{Conclusion and Outlook}

The HERMES RICH detector has been constructed and installed within
13 months after its approval. It has been operating routinely as part of
the HERMES experiment since its installation in May 1998. Its
operation has been stable and reliable for more than two years.
The single photon resolution for ideal tracks is close to the MC predictions.
The particle identification based on the inverse ray tracing technique
has been implemented and its likelihood analysis has been optimized.
More elaborate particle identification schemes are under development. 
Hadron identification by the RICH detector will be a crucial feature 
of the analysis of the current (1998 - 2000) and future HERMES data.

\section{Acknowledgements}

We gratefully acknowledge the DESY management for its support and the DESY 
staff and the staffs of the collaborating institutions for their strong and
enthusiastic support. 
It is a pleasure to acknowledge the assistance of R. Romeo of 
Composite Mirror Applications in
the installation and initial alignment of the mirror array, and
valuable consultations with the SELEX group at FNAL during the design
of the HERMES RICH.
This work was supported by the FWO-Flanders, Belgium;
the INTAS (project 96-274), HCM, and TMR network contributions (ERBFMRXCT960008)
from the European Community:
The German Bundesministerium f\"{u}r Bildung, Wissen\-schaft, Forschung and 
Technologie; the \linebreak Deutscher Akademischer Austauschdienst (DAAD); the Italian 
Istituto Nazionale di Fisica Nucleare (INFN); Monbusho, JSPS, and Toary 
Science Foundation of Japan; the U.S Department of Energy; and the U.S.
National Science Foundation.

\end{document}